\documentclass[12pt]{spieman}  
\usepackage{amsmath,amsfonts,amssymb}
\usepackage{graphicx}
\usepackage{setspace}
\usepackage{tocloft}
\usepackage[numbers]{natbib}
\usepackage{color}
\usepackage[margin=1in]{geometry}
\usepackage[left]{lineno}

\newcommand{\Msun}{\hbox{M$_{\odot}$}}

\newcommand{\micron}{\hbox{$\mu$m}}

\title{Building SPARCS, an Ultraviolet Science CubeSat for Exoplanet Habitability Studies, Technology Advancements, and Mission Training
}

\author[a,b,*]{Evgenya L. Shkolnik}
\author[c]{David R. Ardila}
\author[a]{Logan Jensen}
\author[c]{April D. Jewell}
\author[a]{Tahina Ramiaramanantsoa}
\author[a]{Judd Bowman}
\author[a,b]{Daniel Jacobs}
\author[d]{Paul Scowen}
\author[c]{Christophe Basset}
\author[a]{Johnathan Gamaunt}
\author[e]{Dawn Gregory}
\author[a]{Maria C. Ladwig}
\author[a]{Matthew Kolopanis}
\author[c]{Shouleh Nikzad}
\author[e]{Nathaniel Struebel}
\author[f]{Joe Llama}
\author[g]{Mary Knapp}
\author[h,d]{Sarah Peacock}
\author[a]{Titu Samson}
\author[c]{Mark Swain}

\affil[a]{School of Earth and Space Exploration, Arizona State University, 781 Terrace Mall, Tempe, AZ, USA, 85287}
\affil[b]{Interplanetary Initiative, Arizona State University, 300 E. University Dr., Suite 110, Tempe, AZ, 85281}
\affil[c]{Jet Propulsion Laboratory/California Institute of Technology, 4800 Oak Grove Drive, Pasadena, CA, USA, 91109}
\affil[d]{NASA Goddard Space Flight Center, 8800 Greenbelt Road, Greenbelt, MD, USA, 20771}
\affil[e]{Arizona Space Technologies, 70 S. Val Vista Drive, Suite A3-191, Gilbert, AZ, 85296}
\affil[f]{Lowell Observatory, 1400 W Mars Hill Road, Flagstaff, AZ, USA 86001}
\affil[g]{Haystack Observatory, Massachusetts Institute of Technology, 99 Millstone Road, Westford, MA, USA, 01886}
\affil[h]{University of Maryland, Baltimore County, 1000 Hilltop Circle, Baltimore, MD, 21250}

\cftpagenumbersoff{figure}
\cftpagenumbersoff{table} 
\begin{document} 
\maketitle

\begin{abstract}
The Star-Planet Activity Research CubeSat (SPARCS) is a NASA-funded 6U-CubeSat mission designed to monitor ultraviolet (UV) radiation from low-mass stars. These stars' relatively high-frequency and high-energy UV flares significantly affect the atmospheres of orbiting exoplanets, driving atmospheric loss and altering the conditions for habitability. SPARCS aims to capture time-resolved photometric data in the far-UV and near-UV simultaneously to better characterize the flares and detect the strongest and rarest among them. In addition, SPARCS is testing innovative technology, such as delta-doped detectors with near 100\% internal quantum efficiency and detector-integrated metal-dielectric UV bandpass filters. This mission will increase the technology readiness level of these critical components, positioning them for inclusion in future flagship missions like the Habitable Worlds Observatory. This paper outlines SPARCS’ mission goals and provides an update as the spacecraft is completed and awaits its planned late-2025 launch to a sun-synchronous low-Earth orbit. It also highlights the critical role of small missions in providing training and leadership development opportunities for students and researchers, advancing technology for larger observatories, and shares lessons learned from collaborations between academic, government, and industry partners.
\end{abstract}


{\noindent \footnotesize\textbf{*}Corresponding author,  \linkable{shkolnik@asu.edu} }

\begin{spacing}{2}   

\section{Introduction}

CubeSats have emerged as powerful tools for advancing both space technology and scientific research, particularly in the domain of exoplanetary science and stellar activity monitoring. Historically, large observatories like the Hubble Space Telescope (HST) and the James Webb Space Telescope (JWST) have been the cornerstones of space-based astronomy. However, these flagship missions are costly and often operate with very tight schedules, sharing time among $\sim$1000 programs.

With the advent of more sensitive and powerful detector and other technologies, CubeSats are now a viable vehicle for scientific observations even in the ultraviolet (UV), a historically challenging part of the spectrum because of the need for space-based platforms due to atmospheric absorption, high detector quantum efficiency (QE) in typically photon-starved situations, and stringent contamination control.

By offering lower costs and faster development cycles, CubeSats provide a complementary mode of UV access, enabling long-term monitoring of select targets that would be infeasible with larger observatories. As HST’s UV detectors continue to degrade and interest in UV science grows, missions like SPARCS will help sustain and expand the UV astrophysics frontier.

\section{SPARCS Mission Objectives}

Fifty billion K and M stars (0.1 -- 0.8 \Msun) in our galaxy host at least one small planet in the habitable zone (HZ; e.g., \cite{dres15}). The stellar UV radiation incident upon the planets is strong and highly variable (e.g., \cite{loyd2018,loyd2018b}) and can impact planetary atmospheric composition, habitability, and atmospheric erosion (e.g., \cite{meadows2018,amar22,amar23,amaral2025}). These effects are amplified by the extreme proximity of their HZs to the stars (0.1 -- 0.4 AU). 

In order to characterize the evolution of the high-energy environment of these planets, we are flying the Star-Planet Activity Research CubeSat (SPARCS), a 6U CubeSat (30 cm $\times$ 20 cm $\times$ 10 cm) devoted to monitoring low-mass stars in two UV bands: SPARCS far-UV (FUV: 153 -- 171 nm) and SPARCS near-UV (NUV: 260 -- 300 nm). These bands are more sensitive to the detection of flares than optical wavelengths (e.g., \cite{loyd2018}). In addition, radiation in these ranges strongly impacts planetary atmospheres through  photochemistry (e.g., \cite{tian14,sing2016}) and provides the best diagnostic of unavailable ionizing photons, which govern atmospheric escape through photoevaporation (e.g., \cite{Murray-Clay2009,amaral2025}).

Unlike solar-type stars, K and M stars are known to stay active with high emission levels and frequent flares for most of their lives \cite{shko14,rich19,rich23,loyd2018,loyd2018b,fran2020}.    The UV flux emitted during the super-luminous pre-main sequence phase of low-mass stars, persisting for up to a billion years for the lowest mass M stars (e.g., \cite{schn18}), drives erosion and water loss, as well as photochemical O$_2$ buildup for terrestrial planets within the HZ \cite{luge15,amar22,amar23}. Low-mass stars have FUV to NUV flux ratios $\sim$100--1000 times greater than the Sun \cite{fran16,mile17}, resulting in O$_2$ and O$_3$ levels that are 2--3 orders of magnitude greater for a HZ planet than for a similar planet around a Sun-like star. This could limit the potential for habitability of the planet, while also creating false chemical biosignatures \cite{tian14,doma14,harm15,bixe20}. 

Although several studies have explored the impact of flares on planetary atmospheres (e.g., \cite{chadney2017effect,lee2018effects,fran2020}), the full effects of sustained high levels of stellar activity on planetary atmospheres have not been well studied since UV flare rates and flare energies across stellar evolutionary stages are not well known. 
Given the ubiquity of stellar flares from low-mass stars, 
it is critical to determine the accurate lifetime exposure of their planets to such high radiation, from both quiescent and flare emission levels. Initial efforts are presented by \cite{amaral2025}.

\begin{figure*}[t]
\center
\vspace{-0in}
\hspace*{-.0in}
\includegraphics[trim=0cm 0cm 0cm 0cm,clip=true,angle=0,width=1\textwidth,angle=0]{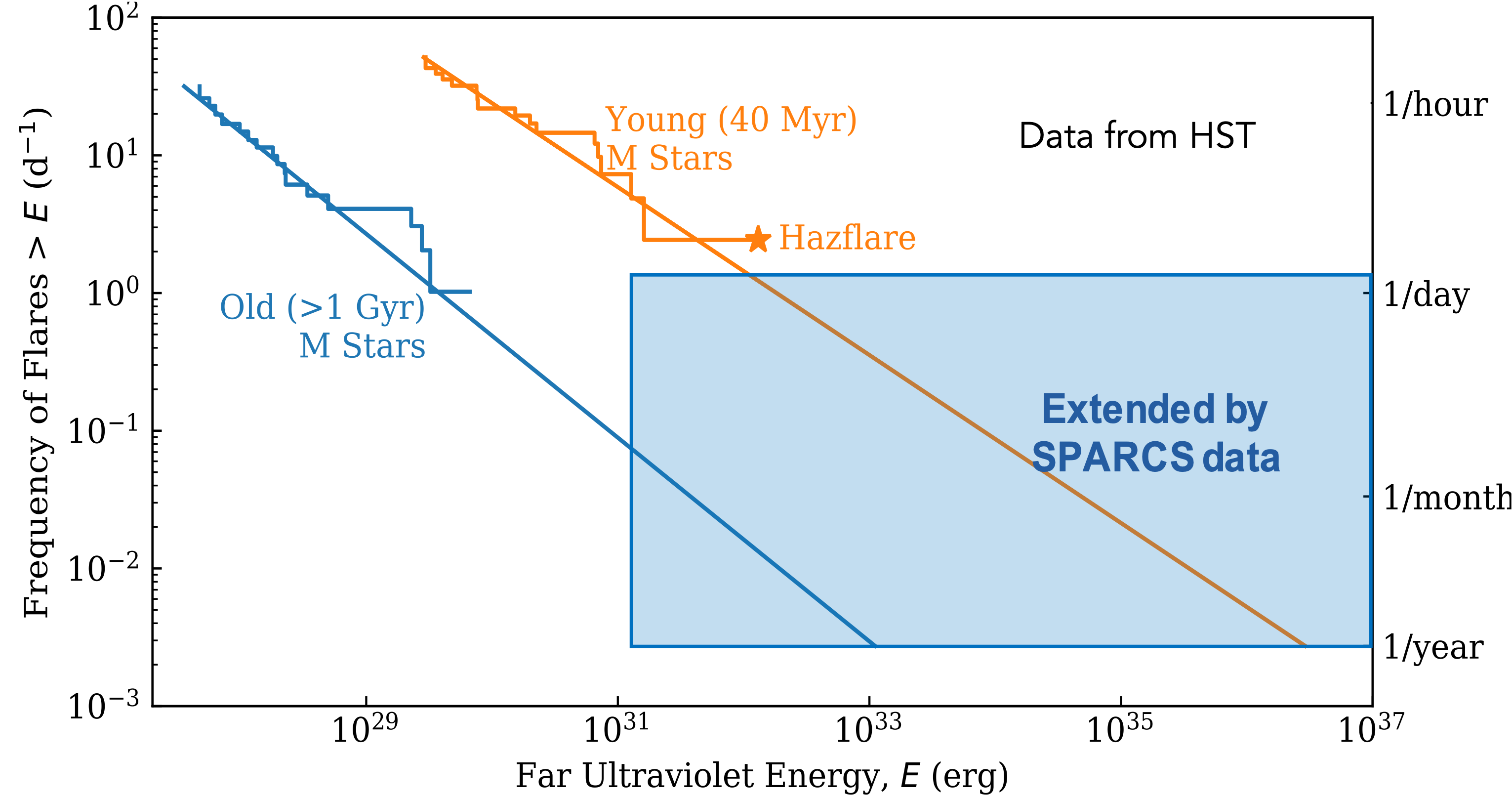}
\vspace{0in}
\caption{Flare frequency distributions (FFDs) for young and old early M stars, adapted from \cite{loyd2018,loyd2018b}. Data from FUV HST observations. SPARCS will extend the FFD from once-per-day flares to once-per-month and and once-per-year flares.
}
\label{fig:ffd}
\vspace{-0in}
\end{figure*}

SPARCS is the first mission dedicated to providing the time-dependent spectral slope (which provides the FUV-to-NUV flux ratio and flare temperature), intensity, and evolution of UV radiation. It focuses on capturing the strongest and rarest flares from a set of low-mass stars and extends this knowledge to the broader population of K and M stars. For this, we must extend our knowledge of the UV time domain from a time scale of tens of hours to months (Fig.~\ref{fig:ffd}). 

SPARCS will operate in a sun-synchronous orbit, allowing near-continuous monitoring of $\approx$20 target stars in both the FUV and NUV bands for 1 -- 3 complete stellar rotations (5 -- 40 days) during its initial one-year mission, enabling the measurement of rotationally modulated quiescence from surface spots and the recording of flares to build accurate flare frequency distributions (FFDs; Fig.~\ref{fig:ffd}).  

Although SPARCS’ primary science objective awaits operations in space, the mission has already contributed to increased engagement with this topic. Through its development, presentations, and early publications, SPARCS has helped draw attention to the importance of UV radiation and stellar flares in shaping exoplanetary environments. Interest in this area has grown substantially in recent years, with many current and upcoming UV missions including low-mass star flare science as a key component.

Since the largest flares are also the rarest, our goal with SPARCS is to increase the cumulative exposure time on low-mass stars in the UV by approximately three orders of magnitude compared to existing archival data from HST and the Galex Evolution Explorer (GALEX). This improvement is not in sensitivity or number of targets, but in time-sampling and total dwell time: SPARCS will observe $\approx$20 stars continuously for 5 -- 40 days each (Table~\ref{tab:target_list}), whereas past UV missions typically obtained only tens of minutes to a few hours per star. This long-duration, high-cadence monitoring over an entire year is critical for detecting rare, high-energy flares and building statistically robust flare frequency distributions (FFD).

\begin{figure*}[t]
\center
\vspace{-0in}
\hspace*{-.0in}
\includegraphics[trim=0cm 0cm 0cm 0cm,clip=true,angle=0,width=1\textwidth,angle=0]{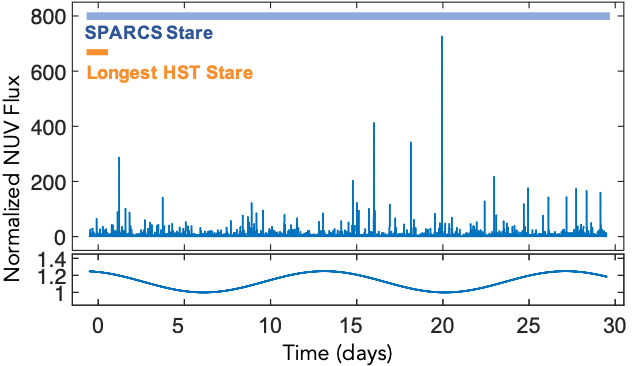}
\vspace{0in}
\caption{Simulated NUV variability of low-mass stars. Left: A 30-day synthetic light curve of a young M0V star in the SPARCS NUV bandpass, showing both flare activity and rotational modulation. The lower panel isolates a 25\% amplitude rotational signal with a simulated 14-day period.
}
\label{fig:lightcurve}
\vspace{-0in}
\end{figure*}

Figure~\ref{fig:lightcurve} (top) shows a simulated NUV light curve of an active, early M star based on an FFD built from HST flare observations (Figure~\ref{fig:ffd}; \cite{loyd2018,loyd2018b}), showing both flare activity and rotational modulation. Figure~\ref{fig:lightcurve} (left, bottom)  shows an example of the quiescent (non-flaring) rotational modulation with an amplitude up to $\approx$25\% \cite{mile17,dosSantos2019} and a simulated 14-day period. With sufficient exposure time and signal-to-noise ratio (SNR $>$ 12), SPARCS is sensitive to this level of quiescent variability. This is supported by the system’s nominal photometric sensitivity, which corresponds to GALEX-equivalent magnitudes of $m_{\mathrm{FUV}} = 18.2$ and $m_{\mathrm{NUV}} = 19.2$ at SNR = 3 in a 10-minute integration. These thresholds define SPARCS' faint-end limits for flare detection and quiescent variability analysis. Despite technology challenges (see Section~\ref{sec:sparcam}), we recover $>$95\% of flares with energies larger than 10$^{31}$ ergs, which is beyond the energies reached with HST observations \cite{rami22a}.

In addition to these science objectives, SPARCS is also advancing two technologies for potential implementation in larger UV missions. Central to this effort is SPARCam, the mission's UV camera system, which employs delta-doped charge-coupled device (CCD) detectors developed at the Jet Propulsion Laboratory (JPL; Section~\ref{sec:sparcam}). These detectors achieve near 100\% internal QE across the UV spectrum incorporating an ultra-thin single crystal silicon with embedded high density of boron atoms in a single atomic layer using molecular beam epitaxy (MBE) \cite{hoenk:1992}. The SPARCam system also incorporates antireflection coatings and a detector-integrated bandpass filter deposited using atomic layer deposition (ALD). The bandpass filter comprises a multilayer metal-dielectric film coating and is optimized for the FUV \cite{hennessy:2015}. Demonstrating these innovations in an operational environment represents a critical step toward their inclusion in future flagship missions like the Habitable Worlds Observatory (HWO). In fact, the development, testing, and delivery of SPARCam has allowed the baselining of delta-doped detectors and detector-integrated metal-dielectric filters for NASA's next MIDEX mission, UVEX \cite{Kulkarni2021,harrison:2023}.

SPARCS is also committed to training the next generation in mission leadership and development. Thus far, the mission has involved more than 11 undergraduate students, 3 Ph.D. students, and 4 postdoctoral scholars, many of whom transitioned to leadership roles in space research and development.  Senior members of the SPARCS team have gone on to lead and contribute to the development of other missions, including several NASA Pioneer, SMEX, MIDEX, Probe, and flagship efforts.

\subsection{Target Stars and Observational Strategy}

Low-mass stars, such as K and M stars, exhibit consistently high levels of UV activity throughout their lifetimes, with 1 -- 2 orders of magnitude higher levels in their first few hundred million years, significantly influencing the atmospheric evolution and habitability of orbiting planets (e.g., \cite{rich23}). SPARCS will observe approximately 20 K and M stars, carefully selected to represent a wide range of stellar ages and corresponding activity levels. These targets will enable a comprehensive analysis of UV radiation's evolution over a star's lifetime and its implications for planetary atmospheres and habitability.

SPARCS addresses critical gaps in our understanding of these stars by performing long-term  simultaneous FUV and NUV monitoring across both quiescent and flare states throughout stellar evolution. The mission employs a single observing mode with two primary observational strategies:

Quiescent Low-Mass Star Measurements: This strategy involves measuring the short-term (minutes) and long-term (weeks) variability and absolute flux of low-mass stars spanning various ages, including young ($\lesssim$ 300 Myr), intermediate-age ($\approx$ 300 Myr -- 1 Gyr), and older stars ($>$~1 Gyr), in both the NUV and FUV, simultaneously.

Flare Stare Measurements: This approach focuses on characterizing flares by measuring their color (FUV-to-NUV flux ratio and flare temperature), energy, frequency, and duration. By monitoring stars over extended periods, SPARCS will refine and extend FFDs, providing detailed insights into the cumulative UV exposure of exoplanets.

To assess SPARCS' ability to detect flares and rotational modulation in the NUV, we performed simulated light curve analyses based on archival HST and GALEX observations of young M-dwarfs, scaled to SPARCS’ bandpass and detector performance. These simulations incorporated realistic stellar flare frequency distributions (FFDs) and amplitudes drawn from \cite{loyd2018}, combined with rotational modulation expected in the UV of spotted low-mass stars \cite{fran16,mile17}. The results informed expected signal-to-noise ratios and sensitivity to both high-energy flares and subtle quiescent variability.

The SPARCS Nominal Target List (Table~\ref{tab:target_list}) reflects the mission’s current best estimate of science targets, focused on late K and early M stars, including well-studied benchmarks like AU Mic and AD Leo. Targets were selected based on a broad range of ages and rotation periods to trace the evolution of stellar UV activity over time, astrophysical relevance (e.g., disk presence, planet hosting), and practical considerations such as UV brightness and field-of-regard accessibility.  The list is labeled ``nominal'' to indicate that selections and observing schedule may shift as commissioning data refine instrument sensitivity and launch and deployment schedules are finalized. SPARCS can also respond to science opportunities to simultaneous observations of targets with JWST or HST. This flexibility ensures SPARCS can maximize its science return within operational realities.

\begin{table}[htbp]
\begin{center}
\label{tab:target_list}
\scriptsize 
\begin{tabular}{|l|l|l|c|c|c|c|c|}
\hline
\textbf{Reason} & \textbf{Name} & \textbf{SpT$^a$} & \textbf{Age (Myr)$^b$} & \textbf{P$_{rot}$ (d)$^c$} & \textbf{FUV ($\mu$Jy)$^d$} & \textbf{NUV ($\mu$Jy)$^d$} & \textbf{Stare Time (d)} \\
\hline
Young & TW Hydrae & M0/K8 & 10$^{1}$ & 3.6$^{2}$ & 2558 & 1.84E+05 & 10 \\
Young with disk & TWA 7 & M2V & 12$^{1}$ & 5.0$^{3}$ & 192 & 1.23E+03 & 10 \\
Young & A0 Men & K4 & 22$^{4}$ & 2.7$^{5}$ & 67 & 8.91E+03 & 10 \\
Young planets, disk host & AU Mic & M1 & 22$^{4}$ & 4.8$^{6}$ & 510 & 3.42E+03 & 10 \\
Young & AF Psc & M4.5 & 22$^{4}$ & 1.1$^{7}$ & 276 & 1.53E+03 & 5 \\
Young planet, disk host & TYC 1766-1431-1 & M1V & 22$^{8}$ & 3.6$^{9}$ & 85 & 8.83E+02 & 18 \\
Young & BD+20 1790 & K5 & 150$^{8}$ & 2.7$^{10}$ & 147 & 1.09E+03 & 10 \\
Young & HIP 106231 & K3 & 150$^{8}$ & 0.4$^{5}$ & 69 & 1.09E+05 & 5 \\
Young & AD Leo & M4V & 25--300$^{11}$ & 2.2$^{12}$ & 480 & 3.54E+03 & 5 \\
Intermediate-age planet, disk host & eps Eri / GJ 144 & K2 & 200--800$^{13}$ & 11.3$^{14}$ & 1820 & 1.60E+07 & 12 \\
Intermediate age & BZ Cet / HIP 13976 & K2.5 & 625$^{15}$ & 9.6$^{16}$ & 166 & 2.63E+05 & 6 \\
Intermediate age & TW PsA / GJ 879 & K4 & 440$^{17}$ & 10.3$^{5}$ & 103 & 3.95E+05 & 10 \\
Intermediate age & HIP 20951 & K0 & 625$^{15}$ & 9.6$^{18}$ & 118 & 1.11E+05 & 12 \\
Intermediate age & HIP 18327 & K0 & 625$^{15}$ & 9.8$^{18}$ & 87 & 8.22E+04 & 10 \\
Intermediate age & HIP 23701 & K2 & 625$^{15}$ & 10.4$^{18}$ & 32 & 5.03E+04 & 10 \\
Intermediate age planet host & GJ 338 B & M0V & 1000--7000$^{19}$ & 16.6$^{19}$ & 122 & 2.89E+03 & 17 \\
Old & sig Dra / GJ 764 & K0 & 5000$^{20}$ & 31$^{21}$ & 180 & 1.23E+06 & 27 \\
Old planet host & Gl 411 & M2V & 5000$^{22}$ & 56$^{23}$ & 81 & 5.04E+03 & 17 \\
Old & GJ 380 & M0/K8 & 5000$^{22}$ & 11.7$^{24}$ & 22 & 9.15E+04 & 12 \\
Old & GJ 825 & M1/M2 & 5000$^{22}$ & 40$^{25}$ & 351 & 3.87E+03 & 40 \\
\hline
\end{tabular}
\end{center}
\caption{SPARCS nominal target list. }
\footnotesize
Notes: 
\newline $^a$Spectral Type. 
\newline $^b$Ages are from young moving group memberships except for the field age stars which we estimate to be roughly 5000 Myr old. 
\newline $^c$Stellar rotation period.
\newline $^d$Quiescent level flux densities predicted in the SPARCS FUV and NUV bandpasses.
\newline \textbf{References:}
$^1$\cite{barr06} \quad
$^2$\cite{dona24} \quad
$^3$\cite{nich21} \quad
$^4$\cite{shko17} \quad
$^5$\cite{mess10} \quad
$^6$\cite{szab21} \quad
$^7$\cite{rams14} \quad
$^8$\cite{bell15} \quad
$^9$\cite{mess17} \quad
$^{10}$\cite{hern15} \quad
$^{11}$\cite{shko09b} \quad
$^{12}$\cite{hunt12} \quad
$^{13}$\cite{mama08} \quad
$^{14}$\cite{frey91} \quad
$^{15}$\cite{perr98} \quad
$^{16}$\cite{batt20} \quad
$^{17}$\cite{mama12} \quad
$^{18}$\cite{nune22} \quad
$^{19}$\cite{gonz20} \quad
$^{20}$\cite{hon24} \quad
$^{21}$\cite{isaa10} \quad
$^{22}$\cite{engl24} \quad
$^{23}$\cite{diez19} \quad
$^{24}$\cite{wrig11} \quad
$^{25}$\cite{byrn89}
\end{table}

\subsection{ Building and Validating} Synthetic EUV--NUV Upper-Atmosphere Stellar Spectra\label{{sec:phoeix}}

Standard stellar atmosphere models (e.g., PHOENIX; \cite{haus97,alla01}) severely under-predict UV emission from low-mass dwarfs (e.g., \cite{peac19a}). These models are intended primarily for comparisons with optical/infrared observations, lack any prescription for the lowest density regions of the upper atmosphere (i.e., the chromosphere, transition region, and corona), and assume local thermodynamic equilibrium (LTE).  	 

We will use the unique time-domain nature of SPARCS observations to validate a new stellar model atmosphere grid, during quiescent and flare states of young and old low-mass stars, suitable for predicting the evolution and variability of the extreme-UV (EUV; 100 -- 1000 \AA) through NUV flux. Developed initially by \cite{peac19a} as part of the HAZMAT program (e.g. \cite{shko14}), the initial spectral models are now available at the Mikulski Archive for Space Telescopes (MAST)\footnote{\url{http://stdatu.stsci.edu/hlsp/hazmat}}
\cite{peac19a,peac19b,peac20} and a first generation grid is in preparation.\footnote{\url{https://science.data.nasa.gov/data-sites/pegasus-stellar-spectra}}  

Since most of the EUV is generally unobservable for low-mass stars due to interstellar medium attenuation, and no EUV-sensitive space telescope exists, these empirically-guided stellar models provide the exoplanetary community with the much-needed input spectra for time-dependent photochemical, climate and atmospheric escape models for a broad range of exoplanets, including rocky planets in the HZ \cite{teal2022}. In turn, new planet atmosphere models calculated under realistic UV conditions provide essential inputs for interpreting transmission and emission spectra obtained with JWST, extremely large ground-based telescopes (ELTs), and future missions, by constraining expected molecular abundances, photochemical pathways, and spectral features under realistic UV irradiation conditions.

We will compute PHOENIX stellar models with prescriptions for the upper atmospheric layers where EUV -- NUV flux originates. FUV and NUV flux originates in the chromosphere and transition region, overlapping heavily with the formation temperatures of the EUV. Ranges of formation temperatures for EUV emission lines (10 -- 91.2 nm) span $\approx$ 30,000 -- 16,000,000 K, temperatures found in the transition region and corona. The EUV continuum forms at cooler temperatures, $\approx$6,000 -- 70,000 K, in the chromosphere and transition region. The FUV-to- NUV spectrum (emission lines and continua, 120 – 300 nm) forms at temperatures between 2,000 -- 200,000 K in the chromosphere and transition region.
The SPARCS FUV and NUV band passes contain emission lines that form at various depths; e.g., Mg II h\&k in the chromosphere at 280 nm (3,500 -- 16,000 K) and He II in upper transition region at 164 nm ($\sim$100,000K). Therefore, these measurements can be used to constrain the temperature structure in the chromosphere and transition region. They do not provide empirical guidance for the corona, so, the models will be unreliable for those specific lines, mostly impacting emission lines from $\approx$10 -- 30 nm.
For an example of similar models guided by GALEX FUV and NUV photometry, see \cite{peac20}.
The quoted EUV flux uncertainties with estimated accuracy to within a factor of two, are computed from inter-model comparisons of synthetic EUV spectra for M and K stars between PHOENIX models, differential emission measure spectra, and Lyman-$\alpha$/EUV empirical scaling relations based on \cite{lins14}.

\subsection{Ancillary UV Science \& Community Participation Enabled by SPARCS}\label{sec:ancillary}

\begin{figure}
\begin{center}
\begin{tabular}{c}
\includegraphics[height=13cm]{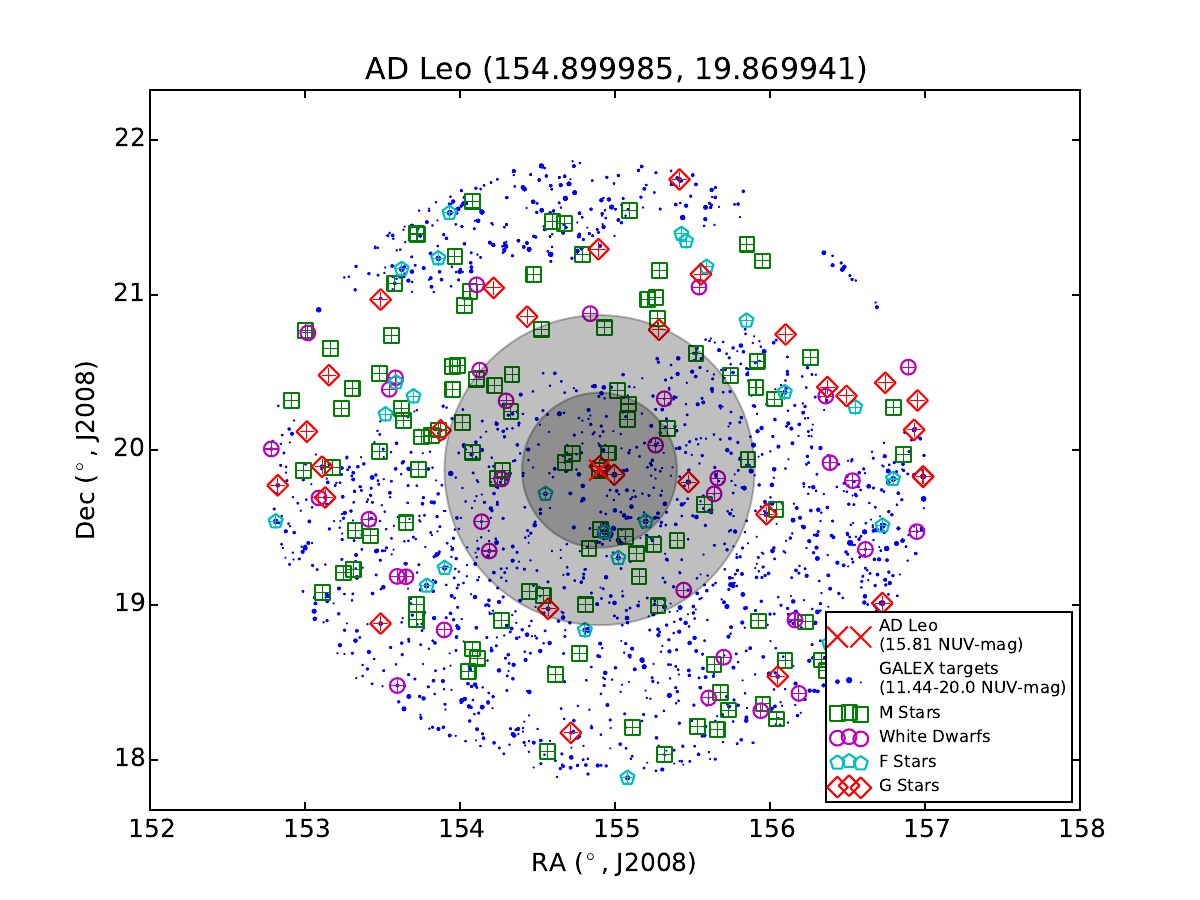}
\end{tabular}
\caption{The SPARCS 40' diameter field of view (FOV; inner dark gray circle) provides opportunities for time-domain UV science for the target star (AD Leo in this case) as well as other objects, including other low-mass stars, white dwarfs for calibration, and ancillary science targets, such as AGNs. The detector will span a larger area than corrected by the optics (lighter gray circle) potentially allowing for observations of additional objects.}
 \label{fig:fov}
 \end{center}
\end{figure}

In addition to its primary targets,  SPARCS enables a wide array of ancillary investigations. Its long-term monitoring (stares of 5 -- 40 days; average of 13 days) with a large 40' field of view (FOV; Figure~\ref{fig:fov}) captures UV variability from other sources.  Given downlink limitations, we extract 10 pre-selected  small subfields (postage stamps) per image centered on the primary target stars and other stars, calibration targets, and active galactic nuclei (AGN). The UV observations of AGN probe mass accretion by the supermassive black hole at the center of its galaxy (e.g., \cite{pete93}). While AGN have been studied by all UV-capable satellites, only a few of those observations were devoted to exploring variability and almost none to short-cadence monitoring (\cite{buhler2024}, and references therein). 
 
There is also interest from the broader community to perform simultaneous ground-based observations with SPARCS UV monitoring of stars to search for correlations among activity indicators. We intend to publish the observing schedule on our website\footnote{\url{https://sparcs.asu.edu/}} during the commissioning phase.  All SPARCS data products, including processed images and light curves of targets, will be archived at MAST.

\section{SPARCam: Technology and Design}\label{sec:sparcam}

\begin{figure}
\begin{center}
\begin{tabular}{c}
\includegraphics[height=10cm]{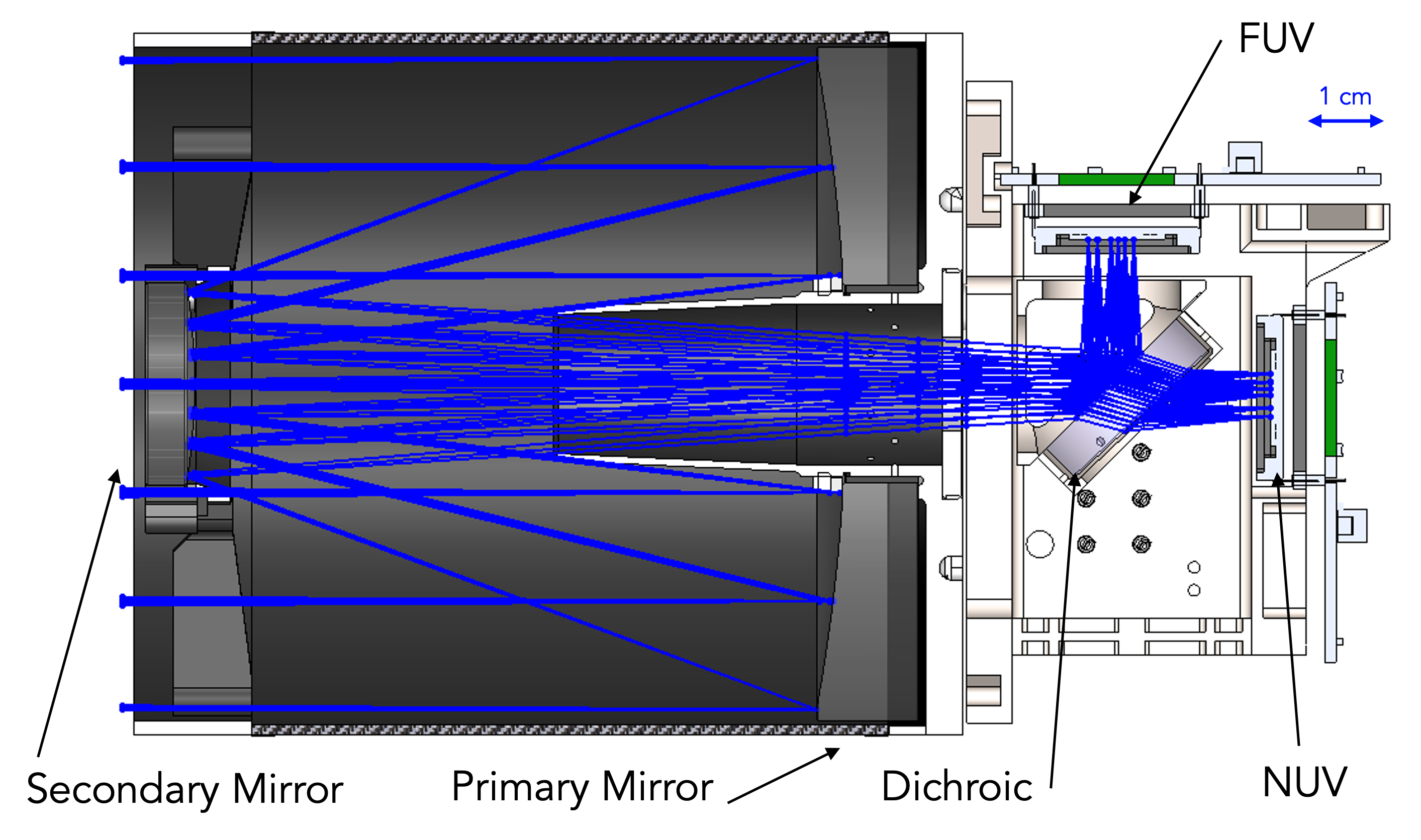}
\end{tabular}
\caption{The SPARCS payload consisting of a telescope with a 9-cm primary mirror and a 3.25 cm secondary mirror feeding a dichroic beam splitter which transmits light to a NUV detector and reflects light to a FUV detector.}
 \label{fig:payload}
 \end{center}
\end{figure}

Light from the optical path is fed to a dichroic beam splitter and directed to NUV (260–300 nm) and FUV (153-171 nm) detectors, enabling simultaneous data collection of stellar UV activity in both bands (Figure~\ref{fig:payload}). SPARCam, SPARCS' highly compact camera system employs two delta-doped CCDs developed at the Jet Propulsion Laboratory (JPL; \cite{hoenk:1992,nikzad17}), one for each UV channel.

SPARCam fits within a 1U volume (10 cm $\times$ 10 cm $\times$ 10 cm), a necessity for 6U CubeSat missions like SPARCS. The camera has a modular design, making it a versatile and volume-efficient system, ideal for CubeSat missions where space and power are limited
\cite{jewell2024}.  

The SPARCam design form and fit are based on the Orbiting Carbon Observatory-3 (OCO-3) Context camera; SPARCam also inherited some of the firmware/data handling  field-programmable-gate-array (FPGA) design from OCO-3 \cite{mckinney2018}. SPARCam uses commercial-off-the-shelf (COTS) components with the exception of the FPGA, which is a space-qualified component. Component selection followed JPL's standard derating practices. Radiation tests were not conducted on the SPARCam hardware, and radiation effects are not specifically mitigated in the system design.

Delta-doping technology enhances detector sensitivity by depositing a thin layer of highly doped epitaxial silicon on the photon-incident surface of the CCD. This process improves the internal  QE to nearly 100\% across the UV spectrum \cite{jewell2024}.
After delta doping, the SPARCS FUV and NUV detectors were optimized individually, with each undergoing specialized backside processing to further increase UV sensitivity in the targeted bandpass.

For the NUV channel, an anti-reflection coating (ARC) of hafnium oxide (HfO$_2$) was applied directly to the detector via ALD, enhancing broadband sensitivity. The downstream side of the dichroic hosts a NUV bandpass filter manufactured by Materion Precision Optics, designed to suppress out-of-band light.

For the FUV channel, a bandpass filter composed of alternating layers of aluminum and aluminum fluoride (AlF$_3$) was directly deposited onto the detector surface using in-vacuum thermal evaporation and ALD. This integrated filter design improves index-of-refraction matching and in-band throughput, providing effective suppression of unwanted visible and red light while maintaining UV sensitivity.

ecause silicon CCD detectors are sensitive to wavelengths up to $\approx$ 1 $\micron$, and low-mass stars are much brighter at red wavelengths than in the UV, careful control of out-of-band (especially red) light is essential. The bandpass filters on both channels provide strong (10$^{-3}$–10$^{-4}$) ``red-leak'' suppression, with the dichroic contributing an additional order of magnitude of out-of-band suppression.

\subsection{Technology Development and Challenges}

The development of SPARCam provided valuable insights into how to build a highly sensitive UV camera system within the volume constraints of a CubeSat. One key takeaway is the importance of modularity and flexibility in camera design. The SPARCam design separates the different functionalities of the camera electronics (e.g., power regulation, data handling, CCD clock bias switching) to different electronics boards, allowing those functionalities to be tested and validated individually during the build. This approach makes for more efficient troubleshooting as problem spots can be quickly isolated by board. Modularity also saves on time and cost if a board replacement or redesign is needed.

To achieve the necessary signal to noise ratio, the CCD detectors are kept cool to reduce the dark-current noise, which requires careful thermal management during both ground testing and integration, as well as during on-orbit operations. The operating temperature of the detectors is set at 238 K (-35 °C) to minimize dark current. Initial designs for the modular system housed the detector on the same board as the analog front-end, clock bias switches, and amplifiers. This layout presented thermal challenges due to the heat dissipation from the proximity electronics. Moving the detectors to separate headboards allows for isolation of the sensors and the ability.
This enables more efficient cooling and allows the system to achieve the low detector temperature and dark current performance required for SPARCS. Thermal testing at JPL and ASU confirmed that the target temperature can be maintained on-orbit. However, the proximity of both heat sources and sinks inside the restrictive payload volume makes for a challenging environment to control and maintain.  The thermal control system uses thermocouples attached to the back of the cold fingers, which are held in contact with the back of the detector packages. Those sensors provide the necessary data to allow the thermal control system to increase or decrease power to the thermo-electric cooler (TEC).  The physical contact between the cold finger and the detector package is fragile and we did encounter intermittent contact during testing.  In addition, interfacing the hot side of the TEC to the spacecraft radiator is not a direct connection and is probably less efficient than it should have been.  Suffice to say that with so many systems in very close contact, managing thermal set points and removing unwanted heat is challenging, yet we have shown it is doable.

\begin{figure}
\begin{center}
\begin{tabular}{c}
\includegraphics[height=10cm]{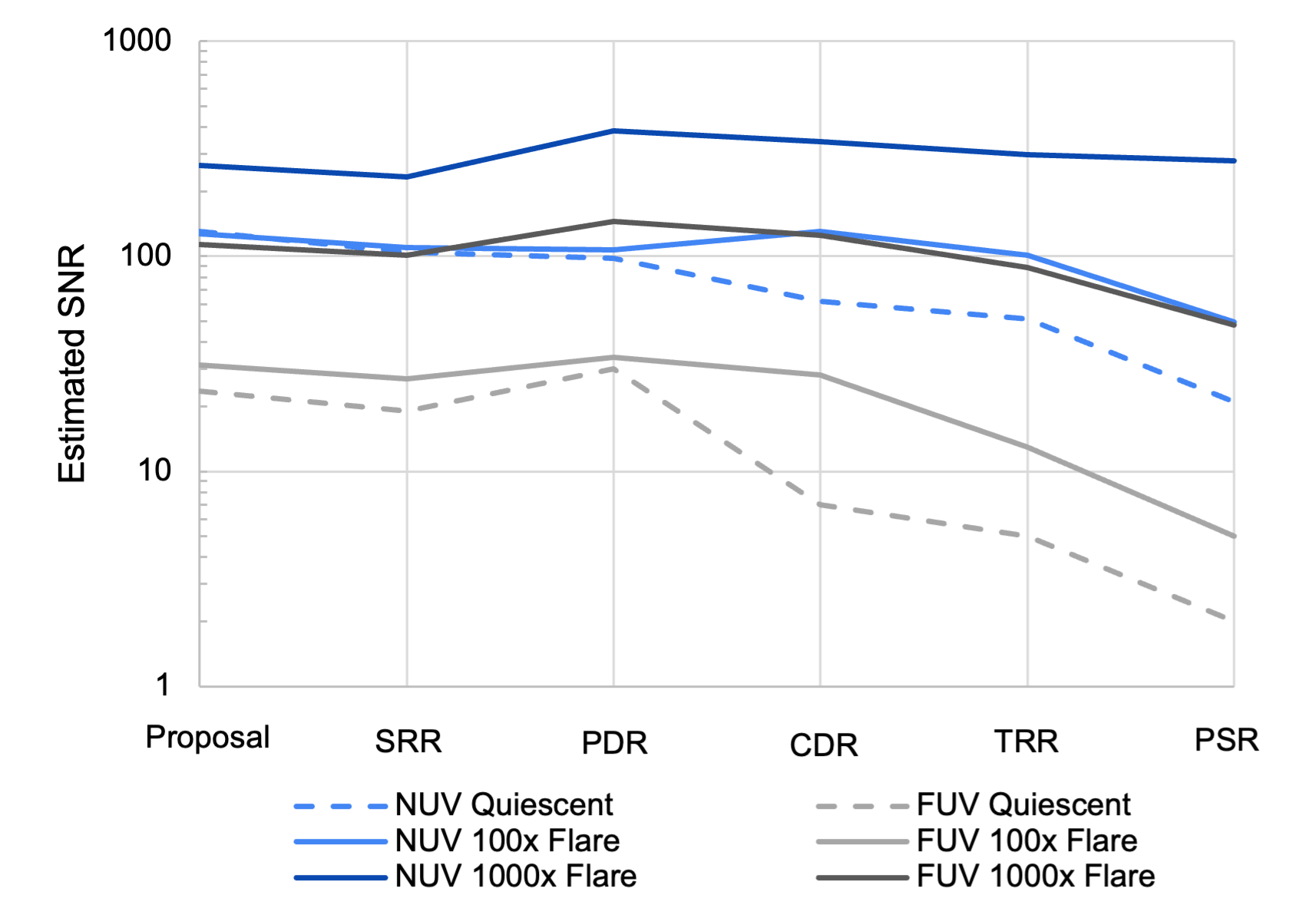}
\end{tabular}
\caption{The expected SNR of both quiescent and example flare observations at different project milestones for target star AD Leo. The quiescent estimates assume a 1-hour observation while the flare estimates assume 30-second observations. The drop in SNR is primarily due to the read noise issue discussed above. (SRR = Systems Requirement Review; PDR = Preliminary Design Review; CDR = Critical Design Review; TRR = Test Readiness Review; PSR = Pre-Ship Review)}
\label{fig:snr}
\end{center}
\end{figure}

Another challenge was related to read noise performance.  Testing of the SPARCam electronics revealed read noise levels more than an order of magnitude higher than the specified requirement. Investigations isolated the source of the added read noise to design errors in the power board. Extensive modifications to the power board—including layout, circuit design, and components, as well as firmware modifications of how the analog output of the CCD was sampled, resulted in a  reduction in the measured read noise to be within $5\times$ the original requirement. Although this is still far from the goal, time and budget did not allow for any further modifications. 
Figure~\ref{fig:snr} shows the decline of the expected  SNR of one of our prime targets, AD Leo, as the project progressed with the bulk of the SNR reduction due to this read noise issue.
Since read noise is not typically a major challenge, it is important to note that relatively high SNR can still be achieved in a CubeSat form factor. For SPARCS, the robustness of the science goals has allowed the mission to preserve its original objectives by shifting from fainter targets to brighter ones.

The SPARCam system highlighted the challenges of working with custom optical components, such as the integrated FUV filter. This included the need for careful matching of the material indices of refraction as well as the associated difficulties of depositing multiple layers of materials onto a detector surface with precise control of thickness and uniformity at the nanometer scale. Ultimately, the success of this approach demonstrates the feasibility of combining detector and filter technologies for compact, high-performance UV cameras. (Further details can be found in \cite{jewell2024}.) Additionaly, the successful integration of delta-doped detectors with specialized bandpass filters marks a significant technological step forward in enabling compact, high-performance UV observations.

\section{Photometric Calibration of SPARCS}

The goal of the SPARCS calibration activities is to provide a dataset that is traceable, repeatable, and useful for a broad range of science measurements. The dataset consists of observing times, count rates, count rate errors, flux densities, flux density errors, along with other measurements such as detector temperatures and source centroid coordinates in the two UV bands for a set of target stars.

SPARCS accuracy requirement is that the flux error should be $\leq$10\%. SPARCS uses the absolute calibration method described in \cite{bohlin:2014}, the same method used in calibration of the photometric modes of GALEX and HST. The counts per second measured on the detector are corrected for out-of-band contributions (i.e., those wavelengths outside of our designed filter bands of 153 -- 171 nm for the FUV and 260 -- 300 nm for the NUV) and transformed into a flux density that corresponds to the photon- and effective area-weighted mean of the source flux over the filter in-band wavelengths. The advantage of this calibration is that it does not require a color correction when applied to different types of objects as the reported flux is not anchored to any specific wavelength.

The quantities involved in the calibration (dark current correction, bias, aperture correction, pixel-to-pixel response non-uniformity, and the system effective area) have been measured on the ground. The role of on-board calibration activities is to monitor changes to the instrument performance to update the measured factors if needed. Complete details on these can be found in \cite{jensen2024} and \cite{ardila2014}.

In orbit, SPARCS will obtain regular observations of white dwarfs (WD) \cite{frymire:2023}, which serve as absolute flux standards to detect and characterize changes in the optical system such as aperture correction. Monitoring changes in instrument throughput and gain between WD observations will be done with other bright point sources in the field.  

 The SPARCS system provides 3 to 4 orders of magnitude suppression at long wavelengths, compared to the peak throughput.  Still, there will be a red leak contribution to the measured flux, particularly for the lowest mass stars. To minimize red-leak contamination, we will apply state-of-the-art spectral models (i.e., \cite{peac19b}) and correct for deviations using pre-flight lab calibrations.

Further details on the SPARCS photometric calibration can be found in \cite{ardila2014}.

\section{Assembly, Integration, and Testing}\label{AIT}

Preparing the SPARCS spacecraft involved meticulous assembly, integration, and testing (AIT) to address the unique challenges of incorporating advanced UV instrumentation into a compact CubeSat platform. The SPARCS team conducted AIT in a custom laboratory at Arizona State University (ASU), designed specifically for the mission's stringent cleanliness requirements as UV optics are highly sensitive to molecular contamination. The cleanroom environment adhered to ISO Class 5 standards. Details about the laboratory setup, requirements, and monitoring see the PhD thesis of Logan Jensen (Jensen 2024, ASU) and \cite{jensen2024}.

The SPARCS optical system includes a 9-cm aperture telescope, the dichroic beam splitter, and the UV detectors, all of which needed precise alignment to ensure optimal performance (Figure~\ref{fig:payload}). The team successfully aligned the SPARCS optics with a series of optical alignment tests and adjustments during payload assembly and integration, ensuring that the system can deliver the required photometric precision once in space. (See details in \cite{jensen2024}.) 

Thermal management was another key focus as CubeSats experience significant temperature variations in low-Earth orbit. SPARCS was subjected to thermal vacuum chamber testing, simulating the harsh space environment to confirm that the payload could maintain stable operational temperatures. This process included thermal cycling tests, which ensured that the detectors and electronics could function within the required temperature ranges without performance degradation. Details of the chamber test environment design and function can be found in Chapter 2 of the PhD Thesis of Johnathan Gamaunt (Gamaunt 2024, ASU) and \cite{gamaunt2022} where details of the chamber and the specific ramping up and down of thermal set points and demonstrated performance of the payload thermal control system are discussed and illustrated.  While we were able to authentically simulate expected conditions on-orbit, making sure the heat flow from the various subsystems in the payload were adequately monitored and controlled required some iteration when some interfaces were found to be less efficient than originally assumed.

Thermal vacuum (TVAC) testing of the SPARCS payload was conducted in a custom-built facility at ASU, tailored specifically for CubeSat-scale instruments. The campaign consisted of three vacuum phases: optical alignment under FUV illumination, functional performance testing, and a ``day-in-the-life'' simulation replicating in-flight conditions. The thermal strategy involved both hot and cold dwells with carefully controlled interfaces to simulate radiator and spacecraft conditions, while maintaining contamination control using an residual gas analyzer (RGA) and a thermal quartz crystal microbalance (TQCM). Notably, the bakeout protocol and interface tuning were iteratively refined based on chamber behavior and unexpected inefficiencies in thermal paths. Lessons learned include the importance of flexible mounting strategies and the sensitivity of TEC-detector interfaces to small variations in contact quality. These insights, particularly relevant to compact platforms like CubeSats, are detailed further in \cite{gamaunt2022}.

 SPARCS uses the Blue Canyon Technologies (BCT) XB-1 bus, featuring  the industry-leading ``XACT'' pointing system, which has been demonstrated to provide $\approx$4'' pointing stability over 10 minutes in similar systems \cite{pong18}.  All of the subsystems in the BCT design have flight heritage. The design includes an S-band radio transmitter suitable for the required data rate, space for the payload radiator, and large fixed solar panels suitable for the sustained payload power draw.  The interfaces to the payload are kept minimal with dual redundant serial data lines and two power zones, one for the payload electronics and one for the bakeout heaters. The purpose of these heaters is to thermally control the optically sensitive/contamination sensitive surfaces to 60$^{\circ}$C for periodic bakeouts on orbit as necessary to maintain throughput performance as determined by regular observations of standard stars.

\begin{figure}
\begin{center}
\begin{tabular}{c}
\includegraphics[height=10cm]{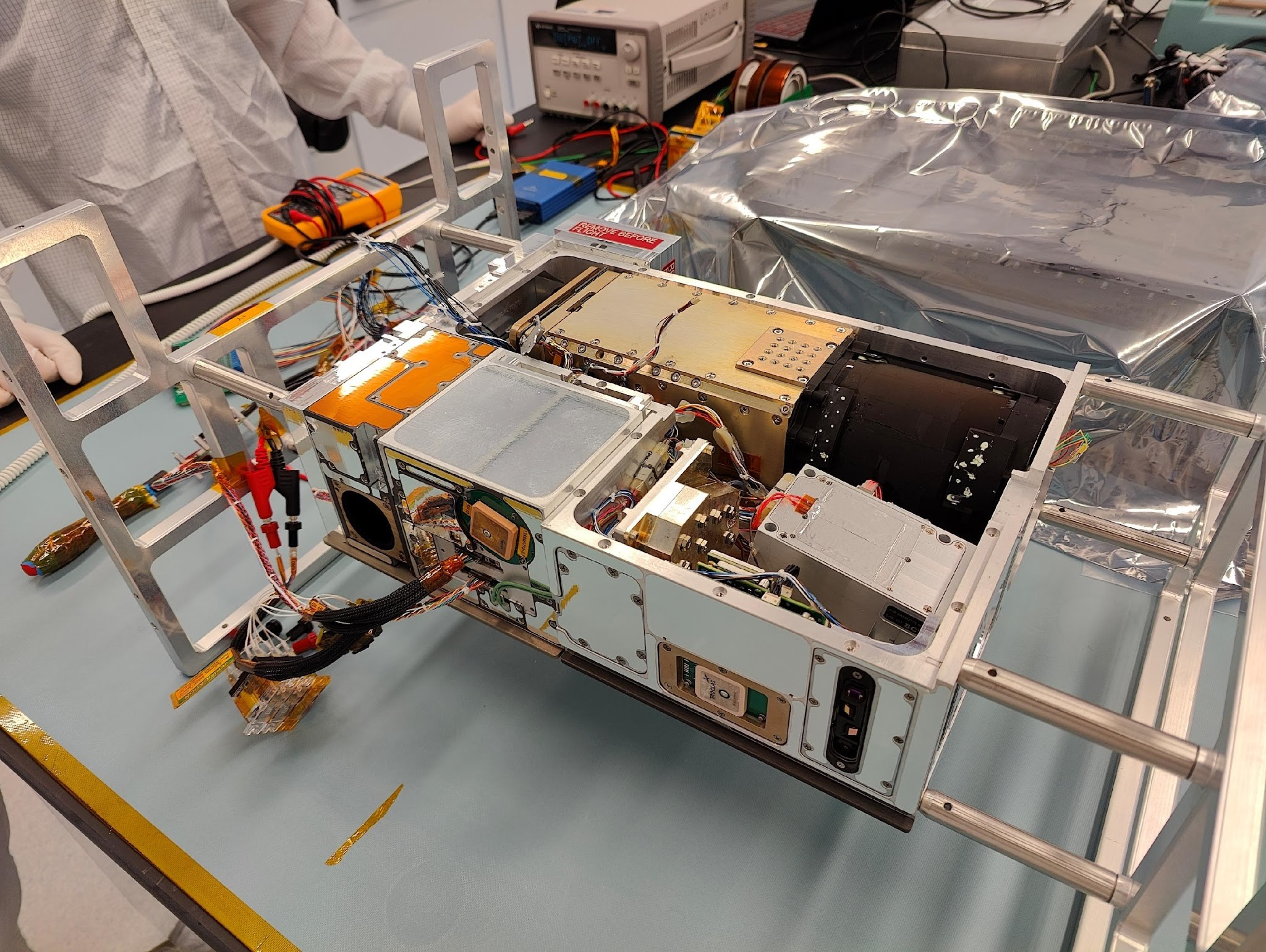}
\end{tabular}
\caption{The SPARCS payload integrated into the spacecraft bus. (Photo credit: Nathaniel Struebel)}
 \label{fig:SPARCS_integrated}
 \end{center}
\end{figure}

\begin{figure}
\begin{center}
\begin{tabular}{c}
\includegraphics[height=10cm]{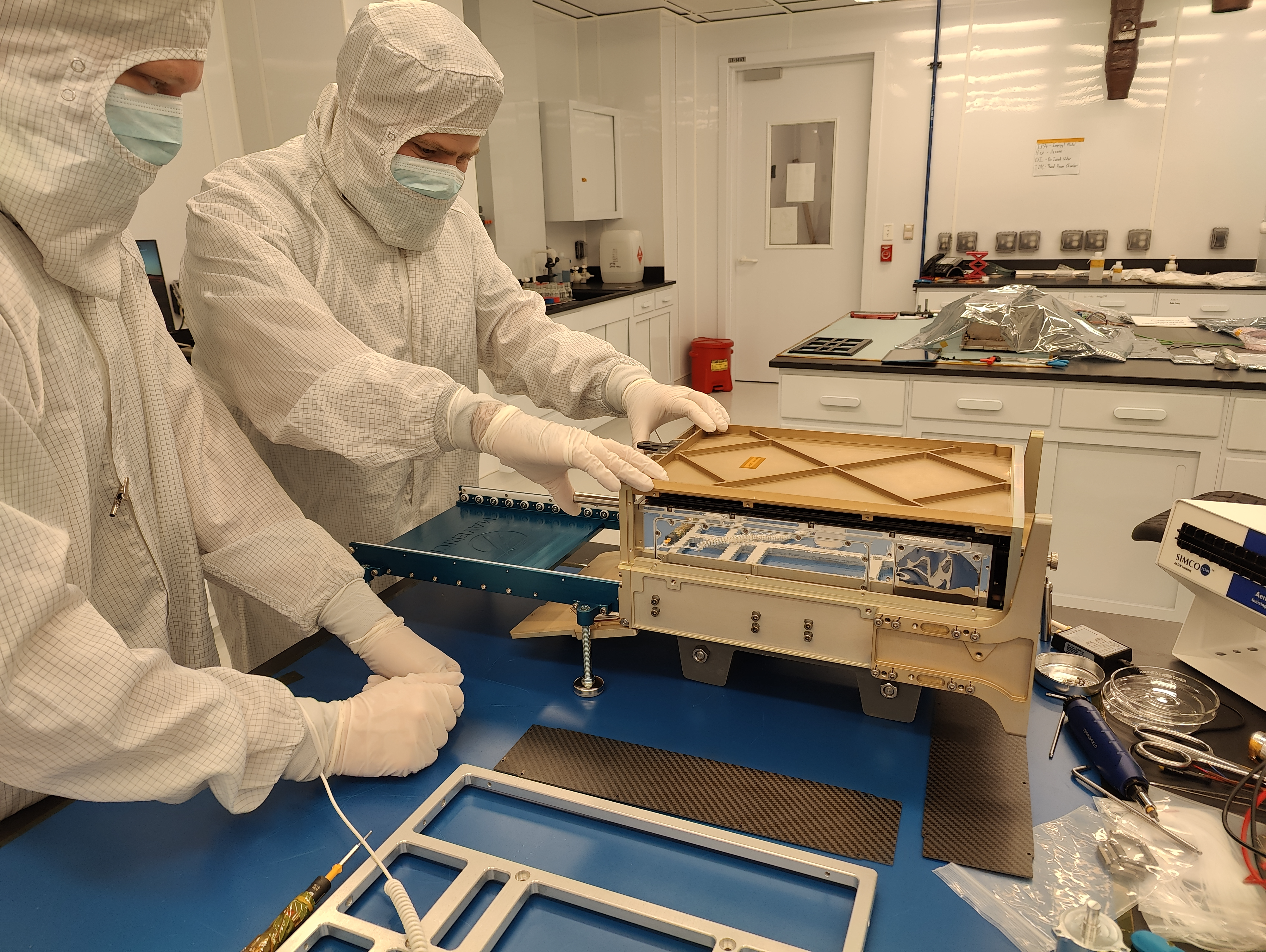}
\end{tabular}
\caption{In preparation for vibration testing, the SPARCS spacecraft was integrated into a Maverick Dispenser. (Photo credit: Ysabella McAuliffe)}
 \label{fig:SPARCS_fit_check}
 \end{center}
\end{figure}

The completed payload was integrated into the bus at ASU and is shown in Figure~\ref{fig:SPARCS_integrated} with the entire spacecraft in its dispenser shown in Figure~\ref{fig:SPARCS_fit_check}.

\section{Software Innovations}\label{software}

The SPARCS mission implemented several innovative software solutions to address the requirements. These developments focused on enhancing reliability, managing dynamic environments, and optimizing performance within the constraints of small spacecraft platforms. The software suite not only supported SPARCS' science objectives but also demonstrated novel approaches that can benefit future missions.

\subsection{Rust-based Payload Software Suite}

The SPARCS science payload uses a custom software suite that manages science observations, performs detector thermal control, and controls command and data transfer flows between the payload and the spacecraft. The payload software suite is fully written in Rust, a modern systems programming language that combines performance comparable to C/C++ with strong guarantees of memory safety and thread safety enforced at compile time \cite{seidel2024}. These features significantly reduce the risk of common programming errors such as null pointer de-referencing, data races, and buffer overflows—critical advantages for autonomous flight software operating in harsh and remote environments. SPARCS is the first to our knowledge to employ a fully Rust-based science payload software suite onboard a space-based astrophysics mission, demonstrating its  viability  for future flight systems.

\subsection{Autonomous Onboard Dynamic Exposure Control}

As discussed above, low-mass stars experience sudden and dramatic increases in brightness during flares. These flares, which can increase a star’s UV brightness by orders of magnitude (e.g., \cite{loyd2018,macgregor21}), pose a significant challenge for fixed-exposure observation systems since very strong flares can reach the non-linear regime of a detector, or even saturate it, losing valuable data. To prevent this, SPARCS employs an autonomous dynamic exposure control system that adjusts the exposure time in near real-time \cite{rami22b}.

A few space-based solar monitoring observatories have used dynamic exposure control based on full-frame statistics that consisted in reducing or increasing the detector exposure time when enough pixels exceed predefined upper or lower threshold values (e.g., \cite{Tsuneta1991,Kano2008,DePontieu2014}). The SPARCS dynamic exposure control system is instead based on individual point source monitoring. It is managed by SPARCS’ onboard payload processor, which performs image reduction to reliably monitor the brightness of the primary target star typically located at the center of the FOV and adjusts the exposure settings in near real-time. The system operates independently for both the NUV and FUV channels, where flare detection in the more sensitive NUV channel stops the longer exposure in the FUV channel to prevent saturation and triggers simultaneous shorter integrations  in both channels.  This allows SPARCS to capture the full range of stellar variability, from quiescent states to extreme flare events, while mitigating detector pixel saturation. 

Since the dynamic exposure control is a feedback system with time delay (the appropriate exposure times can only be applied to the subsequent frame integrations), occasional pixel saturation and blooming\footnote{``Blooming'' refers to the overflow of charge from a saturated pixel to its neighboring pixels in a detector, causing bright artifacts or distortion in the image.} might still happen at the onset of very strong flares. Such occasional bloomed frames are dealt with on the ground by the data reduction pipeline (see Section~\ref{subsec:alice}). The exposure control system is particularly important for capturing flare rise and decay phases, which are critical for understanding the physics of UV flare events.

\subsection{Adaptable ground-based photometric data reduction pipeline}\label{subsec:alice}

So far, every space-based mission dedicated to the photometric monitoring of astrophysical sources had to develop its own custom data reduction pipeline.  The SPARCS ground-based data reduction pipeline, called Adaptive Light Curve Extractor (ALICE; \cite{rami_inprep}), is written in Python and designed to be as generic as possible and easily adaptable to any photometric monitoring observatory. ALICE has thus far been tested on simulated SPARCS science data and actual on-orbit TESS time-series images. Its image reduction module performs dark current correction, flat-fielding, bad pixel correction, as well as cosmic-ray hit identification and correction. Light curve extraction is currently achieved either through circular aperture photometry or a user-defined custom aperture photometry, and backed up with a custom region growing algorithm for autonomous aperture redefinition when blooming is detected, which is useful for imaging systems without anti-blooming features such as the case of the SPARCS detectors.

\section{SPARCS Mission Timeline}\label{sec:timeline}

The SPARCS team completed the payload and spacecraft bus integration and testing in March 2025, and is now awaiting launch in late 2025 via SpaceX's Transporter rideshare program. Post-launch commissioning will take one month, followed by eleven months of primary science operations.

Although SPARCS is designed to operate for at least one year, the absence of consumables on board means that the mission could be extended if the spacecraft remains operational. An extended mission would allow SPARCS to observe additional stars and potentially capture even rarer, higher-energy flare events. All data collected by SPARCS will be made publicly available through MAST, enabling further analysis by the scientific community.

\section{Recommendations and Lessons Learned}\label{lessons}

Science-capable CubeSats and SmallSats have democratized space-based astronomy, enabling broader participation from early-career teams, smaller institutions, and emerging space nations. Their relatively low cost ($\sim$10M -- \$100M) fosters innovation, expands training, and makes ambitious missions more accessible.

SPARCS was the third CubeSat selected by NASA’s Astrophysics Division. Since then, opportunities for small missions have expanded through programs such as the Astrophysics Research and Analysis (APRA) program and the newly-established Pioneers line. With this growth has come a broader and more diverse community of teams navigating the technical, organizational, and cultural challenges unique to small, high-risk missions.

Based on our experience developing SPARCS, we present a set of recommendations and lessons learned that may benefit future missions. Although these lessons are divided into distinct topics, many are rooted in the challenges of communication, whether within the mission team, with collaborators, or with government and industry partners. We begin with insights on internal team dynamics and communication strategies, then broaden to include external collaborations, partner relationships, and structural issues such as mission ownership and funding. We conclude with reflections on financial flexibility and mission resiliency, which synthesize how these factors influence long-term success.

\subsection{Team Integration: Bridging Work Cultures}
SPARCS brought together teams from multiple organizations, some large and some small, and each with its own work culture and management style. Bridging these differences required intentional strategies, such as clarifying roles and establishing shared objectives and expectations. We found that team-building activities, one-to-one relationship building, and in-person meet-ups helped foster open communication and alignment.

To further enhance team integration, we implemented regular knowledge-sharing sessions in the form of full-team meetings, where members presented updates on their work and discussed challenges openly. These sessions not only improved transparency but also built mutual understanding of each partner's priorities and constraints. Creating shared documentation, such as collaborative project roadmaps and schedules, helped ensure everyone remained aligned on key goals and deadlines.

\subsection{Transparent Communication}

Effective communication is key to the success of any mission, especially when working with geographically-distributed teams with diverse institutional objectives. We found it critical that team members feel able to share mistakes or challenges without fear of judgment or repercussions, or else critical issues may remain hidden. 

Cultures where mistakes are hidden are frustrating, but they’re also inefficient. When information is siloed, it can lead to cascading failures, delays, or even mission-threatening errors as problems are discovered late in the development cycle when they are harder and costlier to fix. 

 We prioritized creating an environment where team members felt comfortable sharing mistakes, challenges, or uncertainties without fear of judgment to encourage proactive problem-solving and early intervention on potential issues. To achieve this, we attempted to establish clear communication channels, including regular cross-team meetings and tight project management, allowing for frequent updates, collaborative troubleshooting, and collective decision-making. Equally important were one-on-one check-ins, which provided space for candid discussions and personalized feedback.

By fostering trust and reducing silos, teams are able to address challenges early and maintain a smoother workflow. This culture of transparency not only strengthens collaboration but also improved the mission's efficiency and adaptability. While measuring success in this area can be challenging, the consequences of failure are starkly evident.

\subsection{Collaborative Problem-Solving with Other Missions}
While competition is inevitable during proposal phases, collaboration among funded teams is invaluable during development. Teams often face similar challenges, making peer-mentorship groups or “mastermind groups”, as they are called in business environments, an effective way to share solutions and experiences, typically on at least a monthly basis. NASA’s annual APRA PIs’ meeting is useful for showcasing missions, but a more formalized, regular problem-solving group across all NASA divisions would help mission leaders address common issues efficiently. Participation in such groups should be selective, inviting leaders who value transparency and are willing to share their mistakes and lessons learned for everyone’s benefit.

\subsection{Navigating the Dynamics of Start-Ups and Small Contractors}
Start-ups undergo significant change, especially over the course of a 4 -- 5 year program. The SPARCS timeline intersected with several pivotal moments for its partners, requiring flexibility in adjusting schedules and expectations to accommodate evolving circumstances.
One example, during SPARCS' development, BCT was acquired by the industry giant Raytheon, Inc. This acquisition naturally led to shifts in BCT's priorities, changes in personnel, and increased costs, which necessitated that sections of the quote and relationship be renegotiated, with adjustments to the project timeline. Such transitions underscore the importance of building contingency plans and maintaining open communication to navigate the uncertainties that can arise when working with start-ups and small contractors.

\subsection{Ownership: Challenges of Leading a NASA-funded, but Not-NASA-Owned, Mission}\label{lesson:ownership}

Most large science missions developed in the U.S. are both NASA-funded and NASA-owned, with NASA typically managing regulatory compliance, licensing, and mission oversight. However, smaller missions funded through programs like APRA or Pioneers may be NASA-funded but not NASA-owned. This distinction is not solely based on whether the mission is managed by a non-NASA organization as NASA-owned missions can be managed externally. Rather, the key factor is typically the funding mechanism: missions funded through grants (as opposed to contracts) are generally not NASA-owned, unless the PI is based at a NASA center.

In the case of SPARCS, as an example, securing Federal Communications Commission (FCC) licensing involved additional restrictions on communication frequencies not typically encountered by NASA-owned missions. Furthermore, as of September 2022, the FCC adopted a new deorbit lifetime regulation, which is more stringent for non-NASA-owned missions (5 years vs. 25 years), requiring extra work to ensure compliance or request a waiver. These added layers of complexity highlight the importance of early planning and comprehensive compliance strategies. Recognizing these responsibilities early was critical for SPARCS and provides a useful precedent for future university-led, grant-funded CubeSat efforts.

\subsection{Aligning Funding Profile with Mission Objectives}

The SPARCS mission faced significant challenges due to a funding profile that did not align with the schedule proposed by the team or with the fast-paced development model typically expected of CubeSat-class missions. Although CubeSats are designed for rapid, low-cost execution, often requiring most of the work to be completed within 2–3 years, SPARCS received its funding incrementally through several APRA grants over a longer period, effectively placing the mission on a slower, MIDEX-style timeline. This led to delays, increased costs, and reduced flexibility in project management.

A core issue is that the APRA program, while supportive of CubeSat science missions, does not formally define or structure CubeSat efforts as a distinct class within its funding mechanisms. As a result, CubeSat missions are often subject to the same grant mechanisms and review cycles as less schedule-sensitive science investigations, despite having radically different execution needs. Without up-front or front-loaded funding, teams cannot build momentum, hire dedicated staff efficiently, or navigate critical-path engineering and procurement timelines effectively.
In the case of SPARCS, staggered and misaligned funding created inefficiencies across planning, hiring, and systems engineering, and it compounded other risks, including shifting regulations, partnership changes, and cost growth due to delays.

For future CubeSat missions, especially those funded via APRA, advocating for front-loaded, milestone-based funding structures would reduce risk, enable more efficient execution, and better support the rapid development ethos that makes CubeSats so scientifically and programmatically valuable \cite{shko18}.

\subsection{The Importance of Explicit Cost Reserves}

Cost reserves play a critical role for all mission teams, but especially those attempting to innovate on small budgets. These teams often face unexpected expenses and complexities as they navigate the development of cutting-edge technologies. Explicit cost reserves provide flexibility to address these challenges without compromising mission objectives. For small first-time missions such as SPARCS, the absence of formal financial buffers, which are not typical in science grants (see Section~\ref{lesson:ownership}) but are mandatory in larger missions, significantly increases the risk of delays or scope reductions. Future programs should allow the inclusion of explicit cost reserves to better support emerging teams and raise their chances of success. This is better than the alternative of teams  hiding margin in proposed budgets, which reduces transparency and risks mission failure due to underfunding.

\subsection{Mission Resiliency}

Mission resiliency refers to a project’s ability to adapt to unforeseen challenges while still achieving its core objectives. For CubeSats and other small missions, especially those focused on new technology development and/or training early-career teams, resiliency is not a luxury but a necessity. As discussed above, these missions often operate with tighter budgets, smaller teams, and little to no built-in margin compared to larger programs, making them more vulnerable to setbacks. At the same time, their small scale allows for greater agility and creativity in response.

For SPARCS, one significant challenge was the discovery late in development that the electronic read noise of the camera system was approximately 10$\times$ higher than the original design goal, with modifications eventually reducing this to 5$\times$ higher. Rather than compromise the mission’s science entirely, the team re-evaluated the target list and pivoted to a strategy focused on brighter stars, which allowed SPARCS to maintain sufficient signal-to-noise ratio and preserve its key science deliverables.

Similarly, the contamination potential of any UV system can reduce throughput if not handled correctly. The SPARCS system has been maintained at the required cleanliness levels, but if this had not been the case, we would not have treated it as a failure mode. The calibration pipeline and science margins were designed to absorb such potential losses, allowing the mission to proceed with reduced sensitivity but without loss of core capability.
In this way, mission resiliency can take the form of a system sensitivity that degrades smoothly, without introducing ``cliffs'' (disruptive thresholds) in mission objectives.

\section{Conclusion}

Although SPARCS has yet to begin its primary science operations, the mission has already made meaningful contributions by advancing UV detector technologies, maturing components for future observatories, and training early-career scientists and engineers across all phases of spacecraft development. Many team members have since gone on to contribute to other NASA missions and concepts, carrying forward the skills and experience developed through SPARCS.

Along with the other APRA- and Pioneer-class SmallSats,  SPARCS reflects NASA’s growing support for high-impact science enabled by small, agile platforms. As SPARCS approaches launch, it offers a case study in how CubeSats can support focused scientific goals, technology advancement, and workforce development within a single effort. Should the spacecraft continue operating beyond its nominal one-year mission, SPARCS may provide even greater returns, scientific and otherwise, than initially planned.

With launch approaching, the team now turns its attention to SPARCS’ core science phase as the final step in realizing the mission’s full potential.

\section*{Acknowledgments}

This research is supported primarily by NASA's Astrophysics Research and Analysis Program (APRA; 80NSSC18K0545, 80NSSC21K2069), with some additional support from ASU's Interplanetary Initiative (II) and RadSpace Program, and JPL under a contract with NASA (80NM0018D0004). Contributions by S.P. are supported by NASA under award number 80GSFC24M0006. We are particularly grateful to  Michael Garcia, John Hudeck, and Luis Santos of NASA, Jeffrey Booth and Todd Gaier of JPL, Steve Stem of BCT, and Lindy Elkins-Tanton of the II for all their support and advice.  We also appreciate the efforts of John Hennessy and Robin Rodriguezon for the filters and Nathan Bush for valuable insights on electronics. We are also grateful for the efforts of II staff and students, including Joe DuBois,
Zach Felty,
Huy Dinh,
Sid Vaidy,
Ysabella McAuliffe,
Ishi Shah,
Athul Kodancha,
Tyler Field,
Joe Dukowitz,
Noah Campo,
Taizun Jafri,
Aliyah Webster,
Ritwik Sharma,
and Aaron Bournias.

We also recognize the importance of collaboration with industry partners, Arizona Space Technologies, Hexagon US Federal, BCT , Teledyne E2V, and Acton, whose commitments were instrumental in the mission's development. 

Lastly, we appreciate the thoughtful and supportive comments from the two anonymous referees, who helped improve this manuscript. 

\section{Materials and 
Methods} We acknowledge the use of ChatGPT 4.0 for grammatical and typographical review.

\section{Disclosures}

The authors declare there are no financial interests, commercial affiliations, or other potential conflicts of interest that have influenced the objectivity of this research or the writing of this paper.

\section{Code and Data Availability Statement}

Once SPARCS has flown, the data pipelines discussed in this paper will be made publicly available and the science data and models will be archived at MAST.


\bibliography{refs_master_2024Nov}   
\bibliographystyle{spiejour}   

\vspace{0.5in}
\noindent \textbf{Biographies of authors}

\vspace{2ex}\noindent\textbf{Prof. Evgenya L. Shkolnik} is a professor of astrophysics at ASU’s School of Earth and Space Exploration, Associate Director of the Interplanetary Initiative, and PI of SPARCS. She has spent over 20 years studying exoplanets and the stellar activity of their host stars, from Sun-like stars to the lowest-mass M dwarfs, including ''Sun-as-a-star'' studies and star-planet interactions. She also develops new instrumentation and leads interdisciplinary teams to tackle key questions in astronomy and heliophysics.

\vspace{2ex}\noindent\textbf{Dr. David R. Ardila} is an expert in UV instrumentation and mission formulation in a variety of fields, from planet formation to extragalactic astronomy. He currently serves as Deputy Program Manager for NASA's Exoplanet Program office, at JPL. He served as mission manager for K2, program scientist for CubeSats (at The Aerospace Corporation), NASA-ESA liaison for the Herschel Space Observatory, and instrument scientist for Spitzer's infrared spectrograph. His current interests include advancing NASA's strategic goals in the search and characterization of exoplanets, as well as developing small satellite payloads. 

\vspace{2ex}\noindent\textbf{Dr. Logan Jensen} is an astronomer and systems engineer specializing in small satellite missions and ultraviolet instrumentation. He earned his PhD in Exploration Systems Design from Arizona State University, where he contributed to the assembly, integration, and testing of SPARCS. He is currently the Instrument Systems Engineer on the Landolt mission at George Mason University. His expertise spans contamination control, optical system design, and space mission engineering. 


\vspace{2ex}\noindent\textbf{Dr. Ramiaramanantsoa} is a research scientist at Arizona State University. His areas of expertise include the intrinsic variability of massive stars and the development of software suites for the operation of space-based astrophysics observatories. Dr Ramiaramanantsoa is the lead data scientist for SPARCS mission. He is also a US-based participating scientist for the Ultraviolet Transient Astronomy Satellite (ULTRASAT) mission. 

\vspace{2ex}\noindent\textbf{Prof. Judd D. Bowman} received BS degrees in Physics and Electrical Engineering from Washington University in St. Louis in 1998 and a PhD in Physics from MIT in 2007.  He is director of the Beus Center for Cosmic Foundations at Arizona State University and co-director of the Low Frequency Cosmology Laboratory. 

\vspace{2ex}\noindent\textbf{Prof. Daniel Jacobs} received a BS from New Mexico Tech and a PhD from University of Pennsylvania. He is currently a professor at Arizona State University in the School of Earth and Space Exploration. 

\vspace{2ex}\noindent\textbf{Dr. Paul Scowen} is a Senior Research Astrophysicist at the NASA Goddard Space Flight Center.  His research interests have been focused on a better understanding of both the formation and evolution of massive stars and how they shape the evolution of galaxies.  He also has experience developing ultraviolet instruments and missions to leverage the diagnostic lines in the ultraviolet to gain new insight into these critical objects.



\vspace{2ex}\noindent\textbf{Ms. Dawn Gregory} is an aerospace engineer with over twenty years of experience working on spacecraft of all sizes and complexities.  As a consultant with AZ Space Technologies (AZST), she is currently the systems engineer for SPARCS.  While with AZST she also worked on LunaH-Map and L’TES.  Prior to AZST, Ms. Gregory was a systems engineer at Northrop Grumman and a thermal engineer at Lockheed Martin working on several government and military spacecraft programs.

\vspace{2ex}\noindent\textbf{Ms. Maria Cristy Ladwig} is a Program Manager in the Knowledge Enterprise Program Management Office (KE PMO). She holds a BS in Electrical Engineering and Master’s Degree in Business Administration. She has over 9 years of project management experience in the aerospace industry. She currently serves as PM of NASA’s Star-Planet Activity Research CubeSat (SPARCS). She is a member of the Project Management Institute and PMI certified. 

\vspace{2ex}\noindent\textbf{Dr. Matthew Kolopanis} is a Research Scientist at Arizona State University in the Low Frequency Cosmology Lab.  He is an expert in creating software packages for the use in and development of scientific telescopes and observatories, as well as data reduction and analysis of tera-byte scale data products.  He collaborates with both ground and space based telescope teams to answer questions about the Epoch of Reionization, Cosmic Dawn, and stellar evolution.

\vspace{2ex}\noindent\textbf{Dr. Shouleh Nikzad} is a JPL Fellow, Senior research Scientist, Principal engineer, and the head of the Science Division at NASA’s Jet Propulsion Laboratory. Additionally, she holds visiting faculty and lecturer appointments at California Institute of Technology. She has extensive experience in invention, development, and deployment of devices, detectors, coatings, instrument technologies, and instruments especially in the ultraviolet/optical/near infrared part of the spectrum for space missions and terrestrial applications. She is the camera co-lead for SPARCS and instrument scientist for the MIDEX-UVEX. Her recognitions include being an SPIE Fellow, receiving the SPIE’s Aden and Marjorie Meinel Technology Award, and most recently American Astronomical Society’s Weber Award. She is currently serving as the Chair of the Cosmic Origins Program Analysis Group (COPAG).


\vspace{2ex}\noindent\textbf{Dr. Joe Llama} is an astronomer at Lowell Observatory, Flagstaff, AZ. He is an expert in the detection and characterization of exoplanets using the radial velocity technique. He is also the PI of the Lowell Observatory Solar Telescope, a fiber feed into the EXtreme PREcision Spectrometer where he studies the sun-as-star to better understand the impact of stellar activity on our ability to detect earth-mass planets. 


\vspace{2ex}\noindent\textbf{Dr. Sarah Peacock} is an expert in modeling stellar atmospheres and the high-energy radiation of low-mass stars using the PHOENIX atmosphere code. She has extensive experience computing realistic high-resolution spectra (EUV-IR wavelengths; <0.1 A resolution) of exoplanet host stars, constraining the models with available UV spectroscopic and/or photometric measurements of the targets. These synthetic spectra have been used as input in several studies of exoplanet photochemistry and atmospheric escape. 

\vspace{2ex}\noindent\textbf{Dr. Titu Samson} is an Electrical Engineer at the School of Earth and Space Exploration, Arizona State University. Holding a PhD from Cochin University of Science and Technology, he specializes in advanced RF system design, antenna and PCB development, EMI mitigation, and sensor integration. His expertise includes simulation and testing using AWR, Altium, and LTspice, and he is proficient in Python and MATLAB.

\vspace{2ex}\noindent\textbf{Dr. Swain}, who received his Ph.D. in Physics and Astronomy from the University of Rochester in 1996, is a Senior Research Scientist, Principle Scientist, and Supervisor for the Exoplanet Atmospheres Group at the Jet Propulsion Laboratory.   He is one of the pioneers of detecting molecules in exoplanet atmospheres using near infrared spectroscopy and he has extensive experience in infrared instrumentation. His current research interests include comparative exoplanetology and the characterization of potentially habitable worlds.


\listoffigures

\end{spacing}
\end{document}